\documentclass{article}
\usepackage{graphicx}
\usepackage[english]{babel}
\usepackage{cite}

\title{
\includegraphics[width=0.35\textwidth]{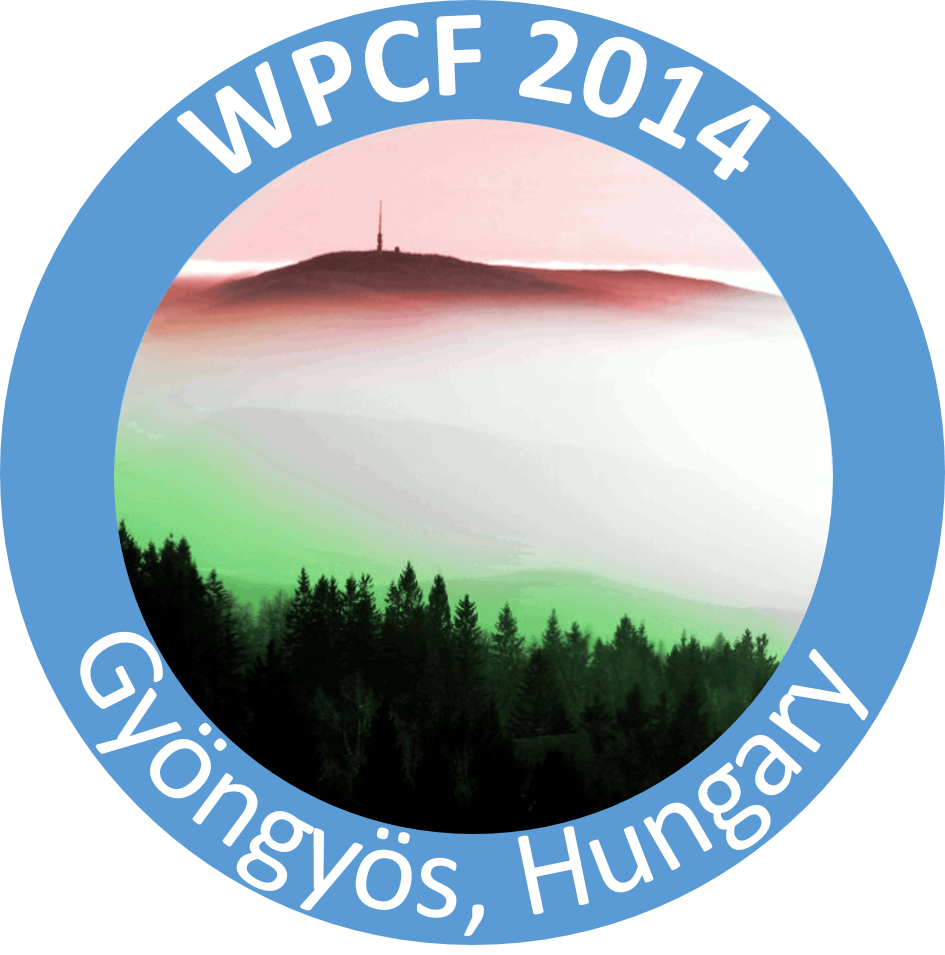}\\[1cm]
Collision Geometry and Flow in Uranium+Uranium Collisions}
\author{{Andy Goldschmidt$^1$, Zhi Qiu$^{1,2}$, Chun Shen$^{1,3}$, Ulrich Heinz$^1$}\\[1ex]
$^1$Department of Physics, The Ohio State University, \\ Columbus, OH 43210, USA\\
$^2$Google Inc., Mountain View, CA 94043, USA \\ 
$^3$Department of Physics, McGill University, \\ Montreal, Quebec H3A 0G4, Canada\\
}

\begin{document}

\fontfamily{lmss}\selectfont
\maketitle

\begin{abstract}
Using event-by-event viscous fluid dynamics to evolve fluctuating initial density profiles from the Monte-Carlo Glauber model for U+U collisions, we report a ``knee''-like structure in the elliptic flow as a function of collision centrality, located near 0.5\% centrality as measured by the final charged multiplicity. This knee is due to the preferential selection of tip-on-tip collision geometries by a high-multiplicity trigger. Such a knee structure is not seen in the STAR data. This rules out the two-component MC-Glauber model for initial energy and entropy production. An enrichment of tip-tip configurations by triggering solely on high-multiplicity in the U+U collisions thus does not work. On the other hand, using the Zero Degree Calorimeters (ZDCs) coupled with event-shape engineering, we identify the selection purity of body-body and tip-tip events in the full-overlap U+U collisions. With additional constraints on the asymmetry of the ZDC signals one can further increases the probability of selecting tip-tip events in U+U collisions. 
\end{abstract}

\section{Introduction}
\label{sec:1}

High energy collisions between heavy ions are used to probe emergent phenomena in Quantum Chromodynamics (QCD), the theory of strong interaction.  One feature of QCD is the transition from hadronic matter to a color-deconfined quark-gluon plasma (QGP) \cite{Heinz:2002gs, Gyulassy:2004vg, Shuryak:2004cy} as the temperature is increased. This transition can occur in heavy-ion collisions of sufficient energy for the system to melt into a hot dense fireball of asymptotically free quarks and gluons.

Relativistic hydrodynamic models have been successful in describing the dynamical evolution of QGP \cite{Kolb:2003dz}. Motivated as a testing ground for these models, a U+U collisions program was recommended in order to study the unique collision geometry resulting from the prolate deformation of the uranium nucleus \cite{Shuryak:1999by,Kolb:2000sd,Heinz:2004ir,Kuhlman:2005ts,Nepali:2006ep, Nepali:2007an,Voloshin:2010ut}. Such a program was carried out in 2012 at the Relativistic Heavy Ion Collider (RHIC) at Brookhaven National Lab \cite{Wang:2014qxa}.

To understand the attraction of uranium, consider that the initial temperature distribution of each QGP droplet is controlled by two main factors: deterministic collision geometry (i.e. the shape of the overlap region between two nuclei), and quantum mechanical fluctuations in the nucleon positions. For spherical nuclei, the collision geometry is entirely a function of the impact parameter.  However, in prolate deformed uranium, the geometry of the initial temperature distribution also depends on the relative spatial orientation of the two nuclei which can be described by the Euler angles between their long major axis.

We focus in this paper on two limiting cases for fully overlapping uranium collisions. In one extreme we have ``tip-tip'' events, defined when the major axes of both nuclei are parallel to the beam direction. The opposite limit are ``body-body'' events, where the major axes of both nuclei are perpendicular to the beam direction and parallel to each other. We are interested in answering the question how, and with what precision, we can distinguish experimentally between these configurations. Their conceptual importance is explained in \cite{Heinz:2004ir}.

\section{The model}
\label{sec:2}

To model the initial energy density distribution of U+U collisions we employ the two-component (wounded nucleon/binary collision) Monte-Carlo Glauber model. We use the deformed Woods-Saxon distribution
\begin{equation}
 \rho(r, \theta, \varphi) = \frac{\rho_0}{1 + e^{(r - r(\theta, \varphi))/d}}
 \label{eq1}
\end{equation} 
to sample the positions of nucleons inside a uranium nucleus.  In Eq.~(\ref{eq1}), the surface diffusiveness parameter is $d=0.44 \mathrm{\ fm}$ and the saturation density parameter is $\rho _0= 0.1660 \mathrm{\ fm}^{-3}$ \cite{Chamon:2002mx, Hirano:2010jg}. The spatial configuration of a uranium nucleus is deformed; we model its radius as \cite{Moller:1993ed}
\begin{equation}
r(\theta ,\varphi ) = r_{ 0 }( 1+\sum_{ l=1 }^{ \infty }{ \sum_{ m=-l }^{ l } \beta_{ lm }Y_{ l }^{ m }(\theta, \phi) }),
\label{eq2}
\end{equation}
where the average radius $r_0=6.86 \mathrm{\ fm}$ is adjusted in such a way that, after folding Eq.~(\ref{eq1}) with the finite charge radius of an individual nucleon, the resulting nuclear charge density distribution agrees with experimental constraints \cite{Hirano:2010jg}. We assume the uranium nucleus is azimuthally symmetric and choose \cite{Filip:2009zz} the non-vanishing deformation parameters $\beta_{20}=0.28$ and $\beta_{40}=0.093$ for the quadrupole and hexadecupole deformations along its main axis, respectively. The choices of these parameters agree well with a recent reanalysis in \cite{Shou:2014eya}, except for $\beta_{20}$ for which Ref.~\cite{Shou:2014eya} gives the value $0.265$.

We use the Woods-Saxon density (\ref{eq1}) to Monte-Carlo sample the nucleon centers and represent each nucleon in the transverse plane by a gaussian areal density distribution about its center:
\begin{equation}
 \rho_n(\vec{\mathbf{r}}_{\perp}) = \frac{1}{(2 \pi B)^{3/2}}e^{-r^2/(2B)} .
 \label{eqn3}
\end{equation}
The width parameter $B = \sigma_{NN}^{in}(\sqrt{s_{NN}})/14.30$ depends on collision energy as described in \cite{Heinz:2011mh}. The sum of these gaussian nucleon density distributions represents the nuclear density distribution for the sampled nucleus at the time of impact and is used to compute the initial energy density distribution generated in the collision. For this calculation, the two-component Monte-Carlo Glauber model weighs a relative contribution from binary collisions $N_b$ and wounded nucleon participants $N_p$ \cite{Bialas:1976ed}.

The binary collision term counts the entropy deposited by pairs of colliding nucleons and is modeled by a gaussian distribution with the same size as a nucleon (see Eq~.(\ref{eqn3})) \cite{Shen:2014vra}; the total binary collision density per unit transverse area is
\begin{equation}
 n_\mathrm{BC} (\vec{\mathbf{r}}_{\perp}) = \sum _{ i,j }{ \gamma_{i,j} \frac{1}{2\pi B}e^{-\left| \vec{ \mathbf{ r}}_{\perp} - \vec{ \mathbf{ R}}_{ i,j }  \right| ^2  / (2B)}}
  \label{eqn4}
\end{equation}
where the sum is over all pairs of colliding nucleons and the normalization $\gamma_{i,j}$ is a $\Gamma$-distributed random variable with unit mean that accounts for multiplicity fluctuations in individual nucleon-nucleon collisions. 

Each struck nucleon is said to be wounded by (or participating in) the collision and contributes a portion of the initial entropy density distributed symmetrically about its center; the resulting total wounded nucleon density per unit area is given by
\begin{equation}
 n_\mathrm{WN} (\vec{\mathbf{r}}_{\perp}) = \sum_{ i }{ \gamma_{i} \frac{1}{2\pi B}e^{-\left| \vec{ \mathbf{ r}}_{\perp} - \vec{ \mathbf{ r}}_{ i, \perp }  \right| ^2  / (2B)}}
 \label{eqn5}
\end{equation}
where the sum is over all wounded nucleons in both nuclei and $\gamma_{i}$ is again a fluctuating normalization factor with unit mean.

We model multiplicity fluctuations in a single nucleon-nucleon collision by taking the normalizations $\gamma_{i,j}$ and $\gamma_{i}$ to be $\Gamma$-distributed random variables with unit mean and variances controlled by parameters $\theta_\mathrm{BC}$ and $\theta_\mathrm{WN}$, respectively. The generic $\Gamma$ distribution with unit mean and scale parameter $\theta$ is given by
\begin{equation} \label{4.1}
\Gamma \left( \gamma ; \theta  \right) = \frac { \gamma^{ 1/\theta - 1 }{ e }^{ - \gamma / \theta } }{ \Gamma \left( 1/\theta  \right) { \theta  }^{ 1/\theta  }  } { , }\quad \gamma \in \left[ 0, \infty  \right)
 \label{eqn6}
\end{equation}
The multiplicity fluctuations from wounded nucleons and binary collisions are related by requiring \cite{Shen:2014vra}:
\begin{equation}
 \theta_{pp} = \frac{ 1-\alpha }{2} \theta_\mathrm{WN}  = \alpha \theta_\mathrm{BC} .
  \label{eqn7}
\end{equation}
where the parameter $\theta_{pp} = 0.9175$ has been fit to the measured multiplicity distributions in p+p collisions \cite{Shen:2014vra}.

The distribution in the transverse plane of the deposited entropy per unit volume is determined by mixing the binary collision and wounded nucleon sources using
\begin{equation}
s_{ 0 }(\vec { \mathbf{r}}_{ \perp  })=\frac { \kappa _{ s } }{ \tau_{ 0 } } \left( \frac { 1-\alpha  }{ 2 } n_\mathrm{WN}(\vec { \mathbf{r} }_{ \perp  })+\alpha n_\mathrm{BC}(\vec { \mathbf{r} }_{ \perp  }) \right) 
\label{eq8}
\end{equation}
where $\tau_0$ is the starting time for the (hydro)dynamical evolution of the collision fireball.
We choose $\kappa_s = 17.16$ and the mixing ratio $\alpha=0.12$ to reproduce the measured charged multiplicities and their dependence on collision centrality in Au+Au collisions at 200 $A$\,GeV. The shape of the resulting energy density distribution in the transverse plane is calculated from the entropy density using the equation of state (EoS) \verb|s95p-v0-PCE| from Lattice QCD \cite{Huovinen:2009yb}.  The initial energy profile is evolved using the viscous relativistic fluid dynamic code package \verb|iEBE-VISHNU| \cite{Shen:2014vra} with specific shear viscosity $\eta/s = 0.08$. Simulations begin at time $\tau_0 = 0.6$\,fm/$c$ and decouple at a temperature $T_{ \mathrm{dec} }= 120$\,MeV. The single particle momentum distribution is then computed using the Cooper-Fyre Formula. A full calculation of charged hadron observables that includes all hadronic resonance decay processes on an event-by-event basis is numerically costly; for this reason we computed only the directly emitted positively charged ``thermal pions'', $\pi^+$, and take this quantity as a measure for total charged multiplicity. At a fixed freeze-out temperature of 120 MeV, the two quantities are related by a constant factor 4.6, $dN_\mathrm{ch}/d\eta \simeq 4.6 \ dN_{\pi^+}/dy$. 

The initial energy density profiles fluctuate from event to event. Each profile can be characterized by the $r^n$-weighted eccentricity coefficients $\varepsilon_n$ and their associated ``participant plane angles'' $\Phi_n$:
\begin{equation}
\mathcal{E}_n := \varepsilon _{ n }e^{ in\Phi _{ n } }=-\frac { \int  d\vec{\mathbf{r}}_{\bot} r^{ n }e^{ in\varphi  }e(\vec{\mathbf{r}}_{\perp}) }{ \int  d\vec{\mathbf{r}}_{\bot} r^{ n }e(\vec{\mathbf{r}}_{\perp}) } ,
\label{eq9}
\end{equation}
where $(r, \varphi)$ are the standard polar coordinates in the transverse plane and $e(\vec{\mathbf{r}}_{\perp})$ is the initial energy density \cite{Alver:2010gr,Gardim:2011xv}.
Through the hydrodynamic evolution, these spatial eccentricities $\{\epsilon_n,\Phi_n\}$ translate themselves into the anisotropic flow coefficients $\{v_n, \Psi_n\}$ \cite{Voloshin:1994mz,Gardim:2011xv,Qiu:2011iv,Qiu:2013wca}:
\begin{equation}
\mathcal{V}_n := v_n e^{in\Psi_n}=\frac{ \int  d \varphi_p dp_T e^{i n \varphi_p} dN/ (p_T dp_T d \varphi_p) }{ \int  d \varphi_p dp_T dN/ (p_T dp_T d \varphi_p)} .
\label{eq10}
\end{equation}

Apart from the Monte-Carlo Glauber model, there exist various other initialization models. These include the IP-Glasma model \cite{Schenke:2012wb}, the MC-KLN model \cite{Kharzeev:2000ph,Drescher:2006ca}, and the TRENTO model \cite{Moreland:2014oya}. As we will see, U+U collisions can provide experimental measurements to distinguish between these various initializations.

\section{Constraining collision geometry with multiplicity, flow, and ZDC cuts}
\label{sec:3}
\subsection{Eccentricity and flow coefficients as a function of multiplicity}

\begin{figure}
  \begin{tabular}{cc}
  \centerline{\includegraphics[width = 0.5\textwidth]{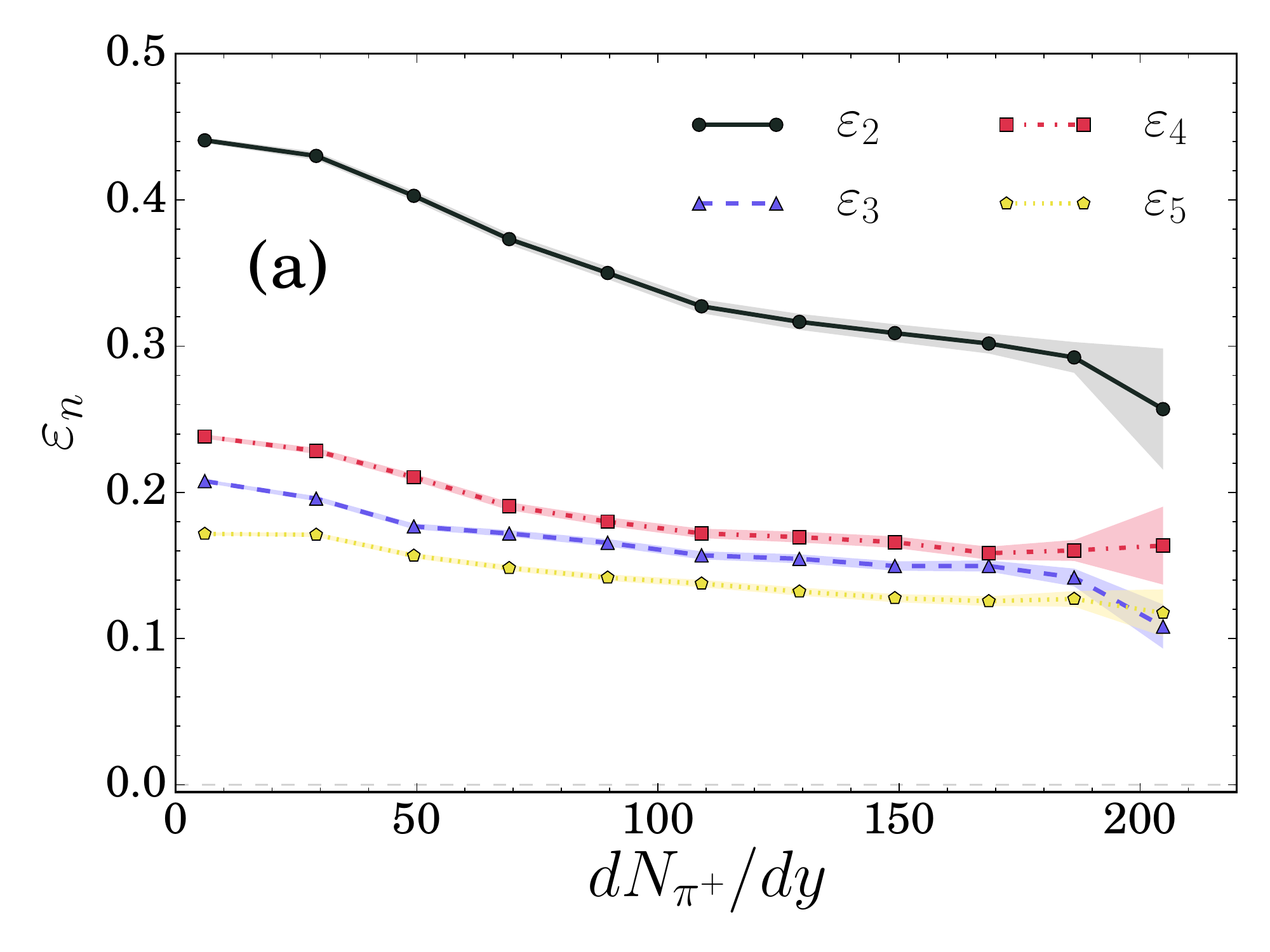}
  \includegraphics[width = 0.5\textwidth]{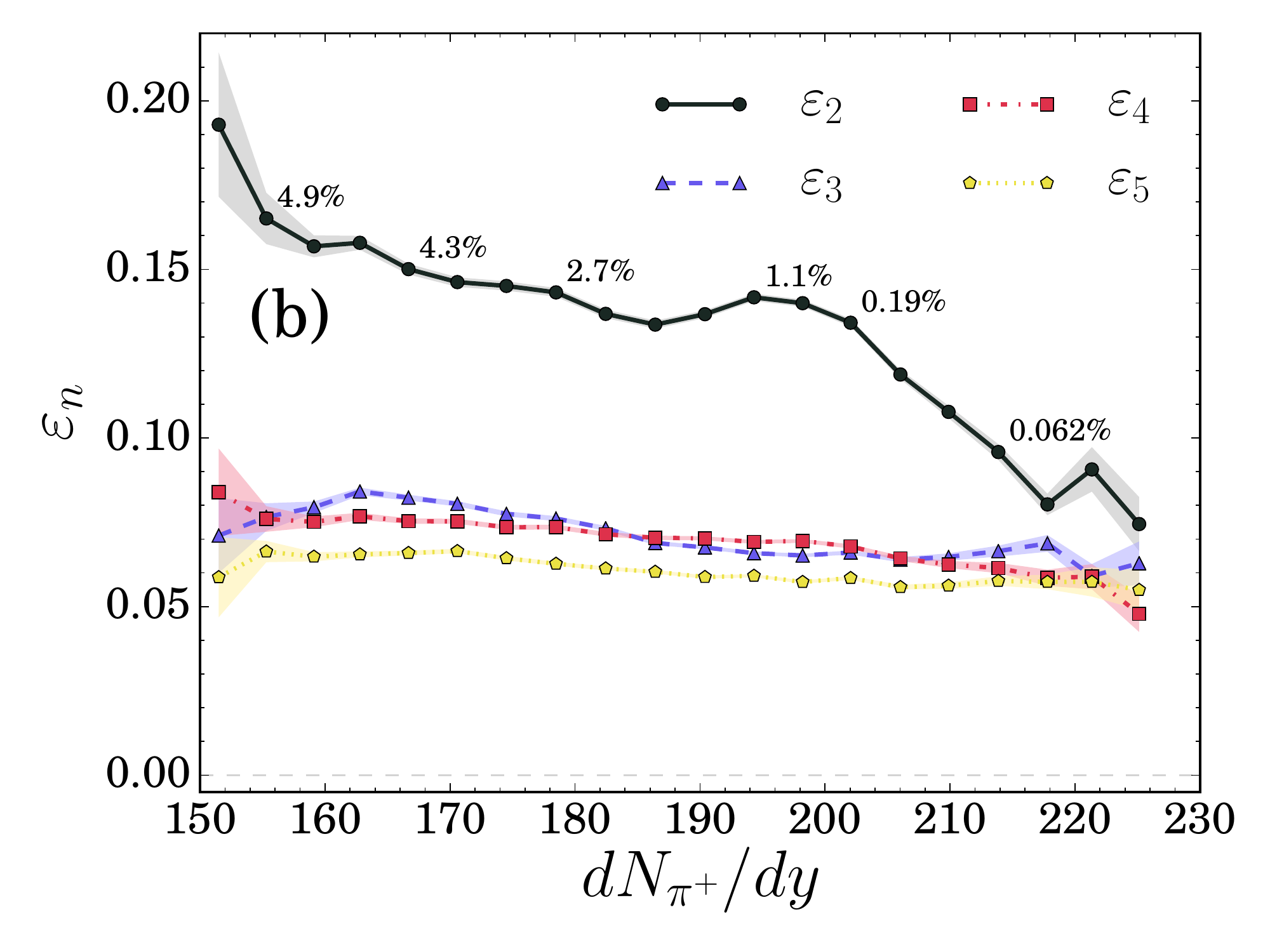}} \\
  \centerline{\includegraphics[width = 0.5\textwidth]{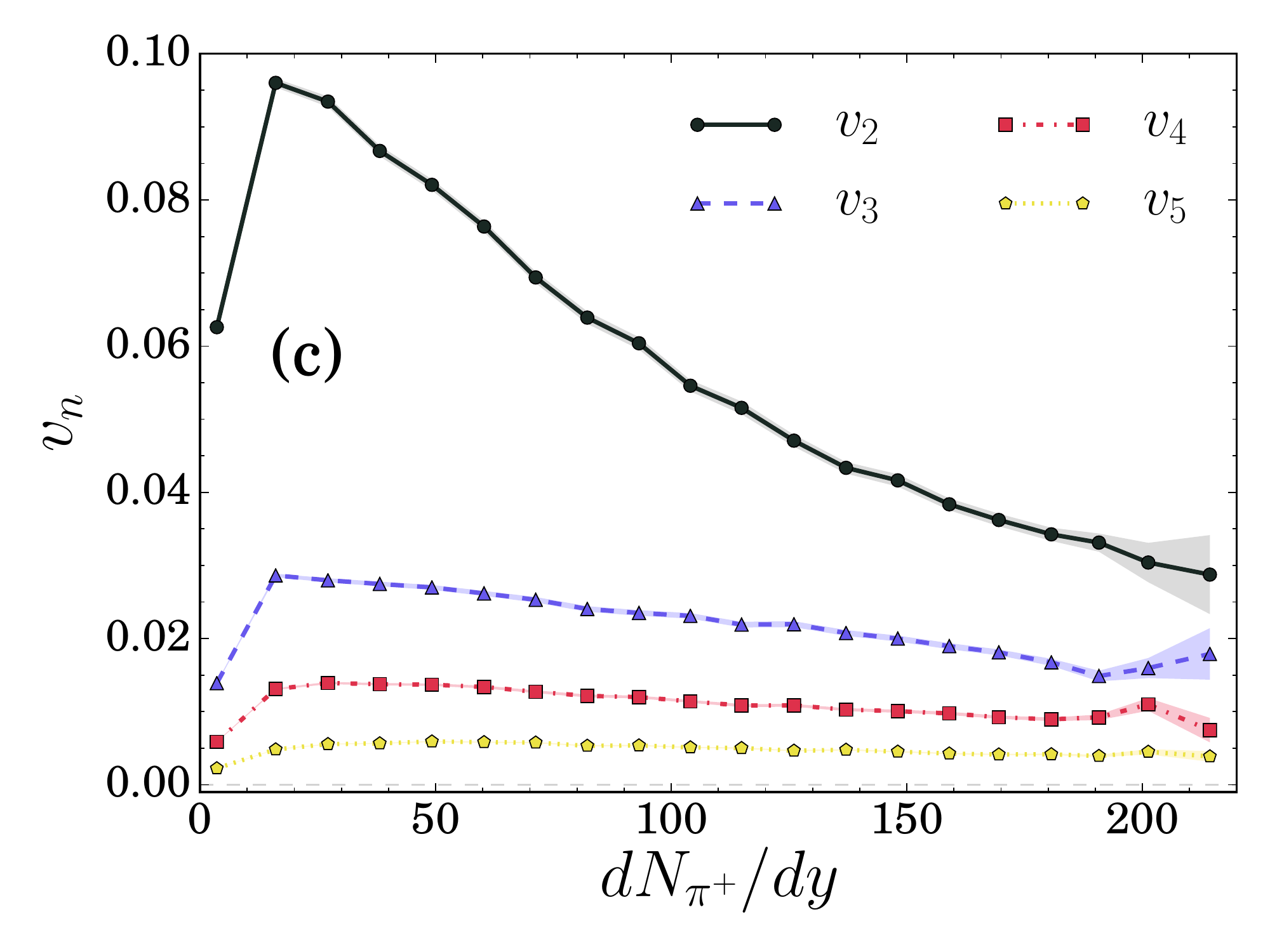}
  \includegraphics[width = 0.5\textwidth]{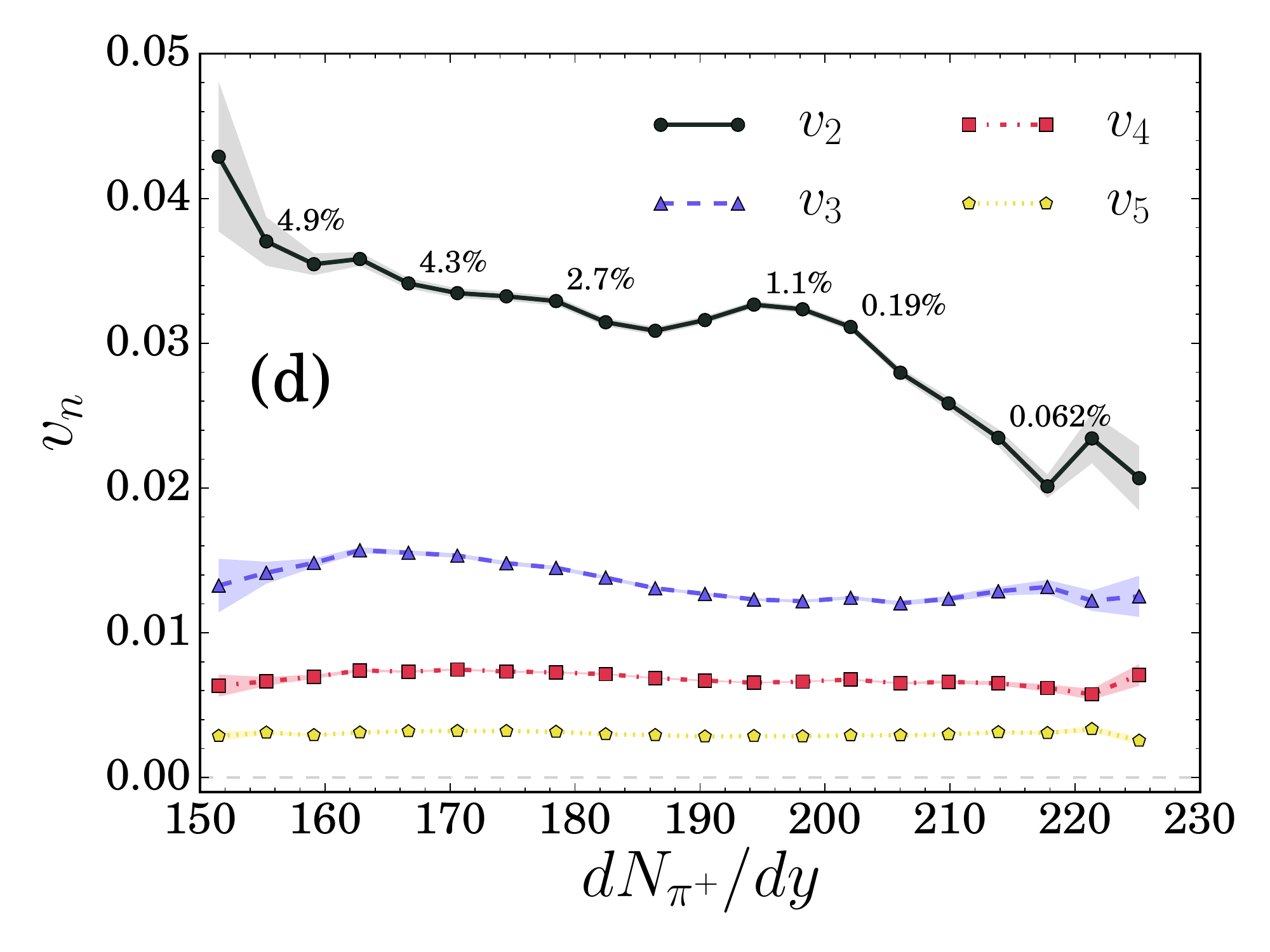}}
  \end{tabular}
  \caption{Panels (a,b) show the event-averaged eccentricities $\varepsilon_n$, before hydrodynamic evolution, panels (c,d) the event-averaged flows $v_n$ after hydrodynamic evolution.  The left panels (a,c) represent 35,000 minimum bias events that include multiplicity fluctuations whereas the right panels (b,d) were obtained from a different set of 35,000 multiplicity-selected events covering the 0-5\% centrality range without multiplicity fluctuations.
  \label{fig2}
  }
\end{figure}

In Fig.~\ref{fig2}, we present the centrality dependence of the initial eccentricities and the final anisotropic flow coefficients of thermal pions for harmonic order $n = 2 - 5$ in U+U collisions at 193 $A$\,GeV. In Figs.~\ref{fig2}a,c minimum bias results are shown as functions of the thermal pion yields, $dN_{\pi^+}/dy$. We notice that the variance of $\varepsilon_{2,4}$ and $v_{2,4}$ in ``most central'' (i.e. highest multiplicity) collisions are larger than in the rest of the centrality range. This is because there the two uranium nuclei are colliding almost centrally (i.e. with impact parameter $b\approx0$) but, as a result of the large spatial deformation, not always with full overlap. A mixture of tip-tip and body-body collisions in these high-multiplicity events increases the variance of the initial $\varepsilon_{2,4}$ which then drives a larger variance in $v_{2,4}$. 

In Figs.~\ref{fig2}b and \ref{fig2}d, we increase the statistics and focus on the 0-5\% most central U+U collisions. We find a ``knee'' structure in the high multiplicity regime ($<0.5\%$ centrality) for both $\varepsilon_2$ and $v_2$. This can be understood as follows: First, while the ellipticity in the transverse plane for a tip-tip collision is small (as the overlap area is approximately circular), body-body collisions produce an ellipsoidally deformed overlap region with larger ellipticity $\varepsilon_2$. Second, although fully overlapping tip-tip and body-body collisions share the same number of participants, more binary collisions between nucleons can happen in the optically thicker tip-tip event, implying (in our two-component Glauber model) a larger initial $dS/dy$ deposited in the tip-tip configuration. In the presence of fluctuations which lead to a range of $\varepsilon_2$ values for a given $dS/dy$ and vice-versa, the larger average multiplicity in tip-tip collisions implies an increasing bias toward small $\varepsilon_2$ when selecting events with larger and larger values of $dS/dy$. This preferential selection of tip-tip orientations at high multiplicities accounts for the appearance of a knee structure in the initial ellipticity \cite{Voloshin:2010ut} (Fig.~\ref{fig2}). We see in Fig.~\ref{fig2}c that the knee is preserved after an event-by-event hydrodynamic simulation when plotting the elliptic flow of the final particle distribution as a function of multiplicity. 

We emphasize that experimental results from STAR do not show this knee structure \cite{Wang:2014qxa}. Considering the preservation of the structure after hydrodynamic evolution as seen in Fig.~\ref{fig2}, we conclude that, in contrast to Au+Au collisions where it has been extensively tested, the two-component MC-Glauber model fails to correctly identify entropy production in ultra central U+U collisions where the knee is predicted by the model but not found experimentally. Hence the non-linear dependence of multiplicity on the number of wounded nucleons observed in spherical Au+Au and Pb+Pb collisions as a function of collision centrality cannot be attributed to a binary collision component as implemented in the two-component MC-Glauber model. 

\begin{figure}
\centering
  \includegraphics[width = 0.7\textwidth]{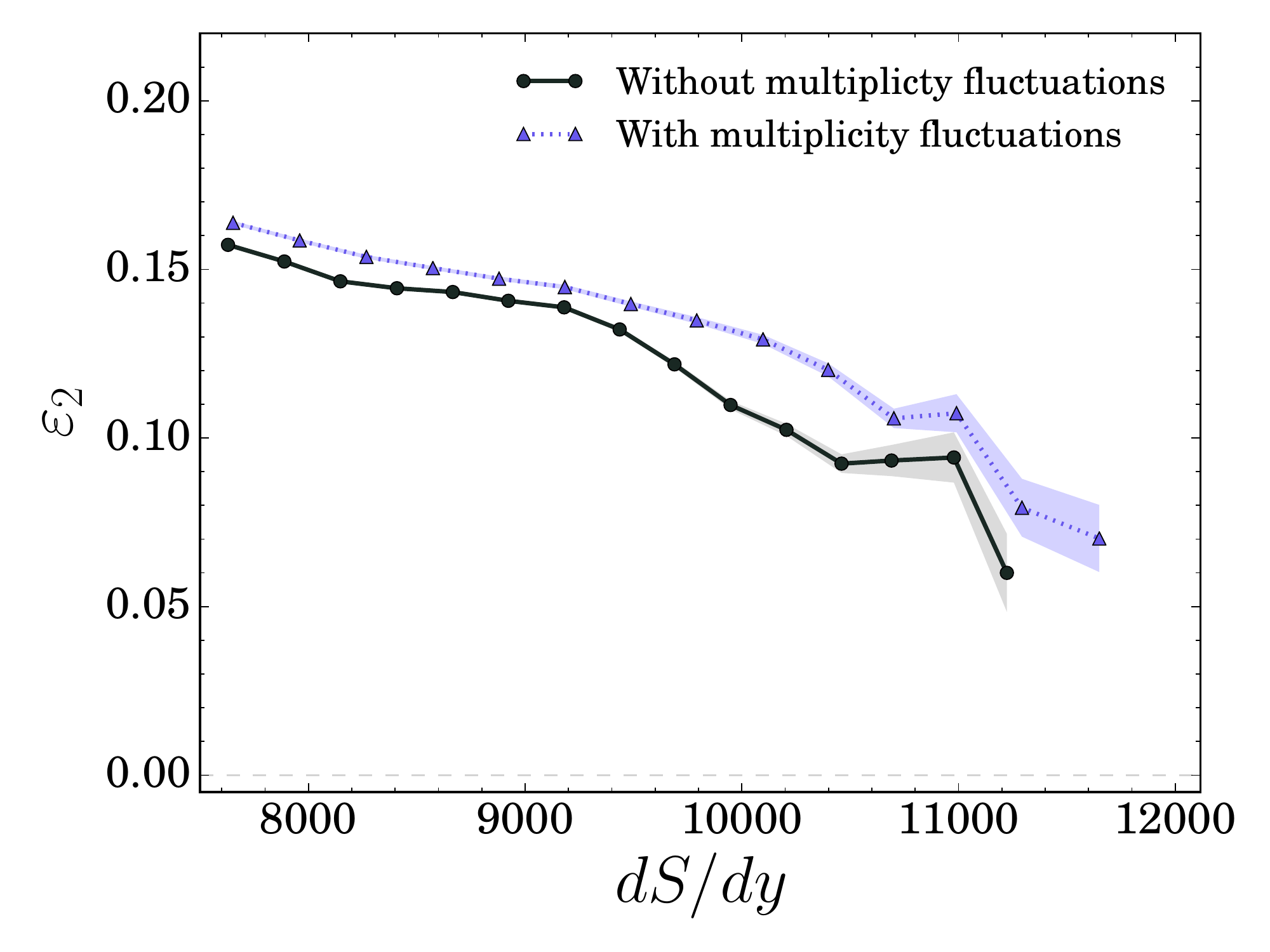}
  \caption{The ellipticity $\varepsilon_2$ as a function of $dS/dy$ from the MC-Glauber model, for collisions roughly in the 0-5\% centrality range, with (blue dashed line) and without (black solid line) including multiplicity fluctuations from single p+p collisions.
  \label{fig:3}
  }
\end{figure}

Some corrections to the entropy production in these ultra central events arise from the inclusion of p+p multiplicity fluctuations.  We see in Fig.~\ref{fig:3} that adding said multiplicity fluctuations weakens but does not erase the knee structure in $\varepsilon_2$ vs. $dS/dy$. Hence, this effect alone does not appear sufficient to reach agreement of the MC-Glauber model with data for ultra central U+U collisions.  We acknowledge that more drastic fluctuation models \cite{Rybczynski:2012av} have been suggested in order to more successfully adjust the theoretical predictions of MC-Glauber to experimental results.

Also of note is the success of the gluon saturation physics as implemented in the IP-Glasma model. Interestingly, this model is able to simultaneously accomodate a strong nonlinearity of $dN/dy$ as a function of $N_\mathrm{part}$ in Au+Au and Pb+Pb and a weak dependence of $dN/dy$ on collision orientation in central U+U at fixed number of participants, while the MC-Glauber model cannot \cite{Schenke:2014tga}. Alternatively, it has been suggested in \cite{Adler:2013aqf} that a model that produces entropy according to the number of wounded valence gluons (rather than wounded nucleons) can also reproduce the observed {\it nonlinearity} of $dN/dy$ as a function of particpant nucleons in Au+Au and Pb+Pb at RHIC and LHC, without a binary collision component. It would be interesting to study the preduction of such a model for central U+U collisions of varying orientations.

\subsection{Selecting high overlap events with combined ZDC and multiplicity cuts}

In an experimental analysis of relativistic heavy ion collisions, the charged hadron multiplicity, $dN_\mathrm{ch}/dy$ and its elliptic flow coefficient $v_2$ can be used to classify events. Hydrodynamic studies have shown that the initial $\varepsilon_2$ maps linearly to the $v_2$ of hadrons \cite{Qiu:2011iv} and the the initial $dS/dy$ is monotonically related the final total particle multiplicity, $dN/dy$ \cite{Shen:2014vra}. We can therefore use $dS/dy$ and $\varepsilon_2$ from the initial conditions as a satisfactory proxy for charged hadron $dN_\mathrm{ch}/dy$ and $v_2$ to test whether we can select the fully overlapping tip-tip and body-body U+U collisions. 


For our analysis, we make theoretical approximations for the use of experimental forward and backward zero degree calorimeters (ZDCs). Placed at zero degrees far from the colliding pair, ZDCs catch information about the spectator neutrons that pass through a collision without participating. We classify our collisions by using the number of spectators $N_s = 476 - N_\mathrm{part}$ to mimic the experimental ZDC signal \cite{Kuhlman:2005ts}. For our study we look at 65,000 events in the $1\%$ most participating ZDC range ($N_s < 19$).  Selecting on the most participating ZDC collisions allows for a restriction of the set of collisions to more fully overlapping events. In such a regime, any initial geometric effects should come more exclusively from the deformed shape of the uranium nucleus.

We define the tip-tip and body-body event classes using the pair of angles $(\theta_{1,2}, \phi_{1,2})$ from the two incoming nuclei, where $\theta$ denote the polar angle between the long major axis of the uranium nucleus and the beam direction and $\phi$ is the azimuthal angle in the transverse plane. An event is defined as tip-tip if $\sqrt{ \cos^2 \theta_1 + \cos^2 \theta_2 } > 0.86$. We classify an event as a body-body event if both $\sqrt{\cos^2 \theta_1 + \cos^2 \theta_2 } < .31$ and $\left|\phi_1 - \phi_2 \right| < \pi/10$.  The polar angle constraints imply that for equal $\theta_1 = \theta_2 $, this common angle $\theta$ is less than $\pi/10$ for tip-tip and greater than $4\pi/10$ for body-body.  For body-body events, the additional azimuthal constraint forces alignment of the long major axes.

\begin{figure*}
\begin{tabular}{cc}
 \centerline{\includegraphics[width = 0.5\textwidth]{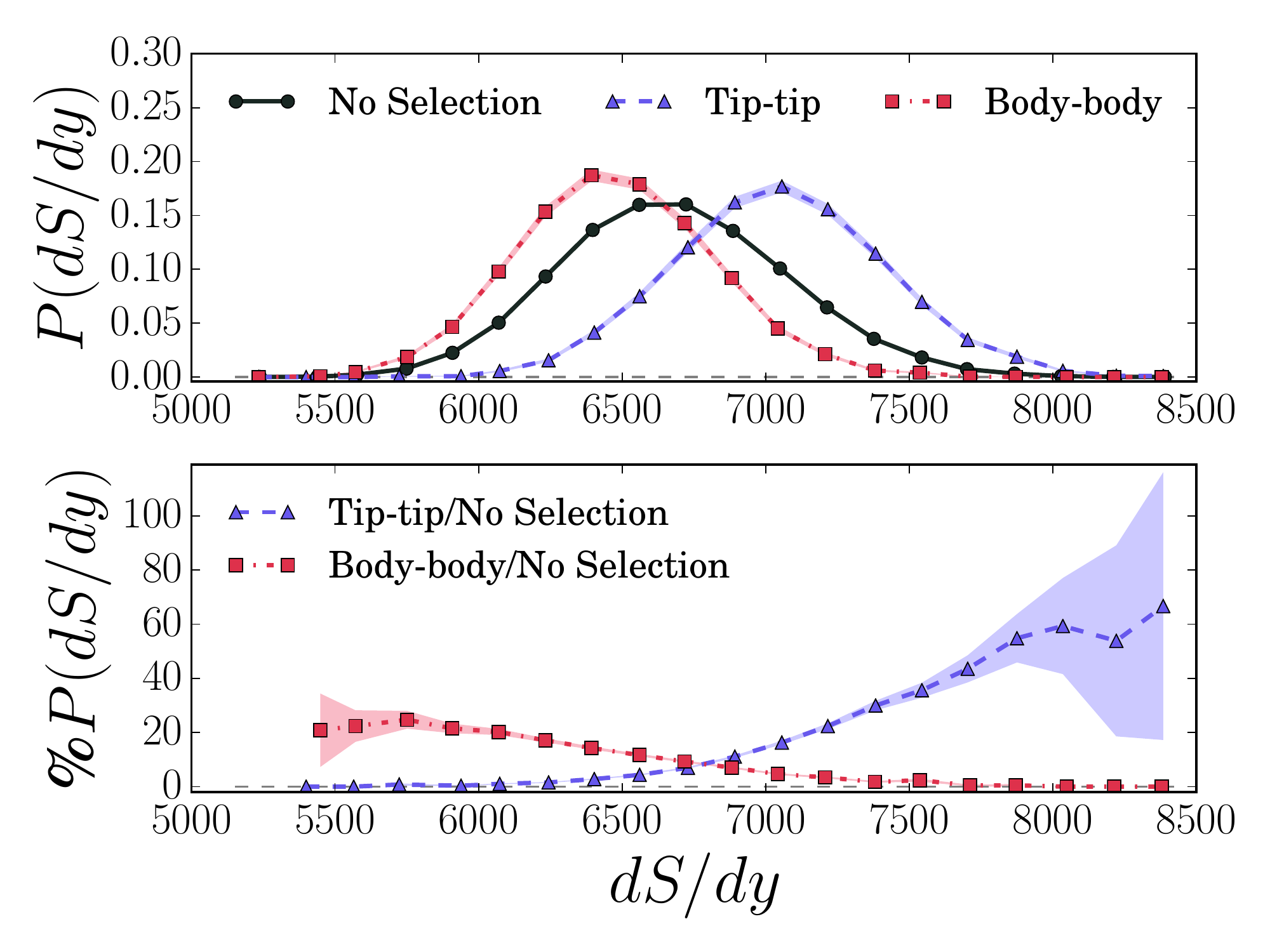}
 \includegraphics[width = 0.5\textwidth]{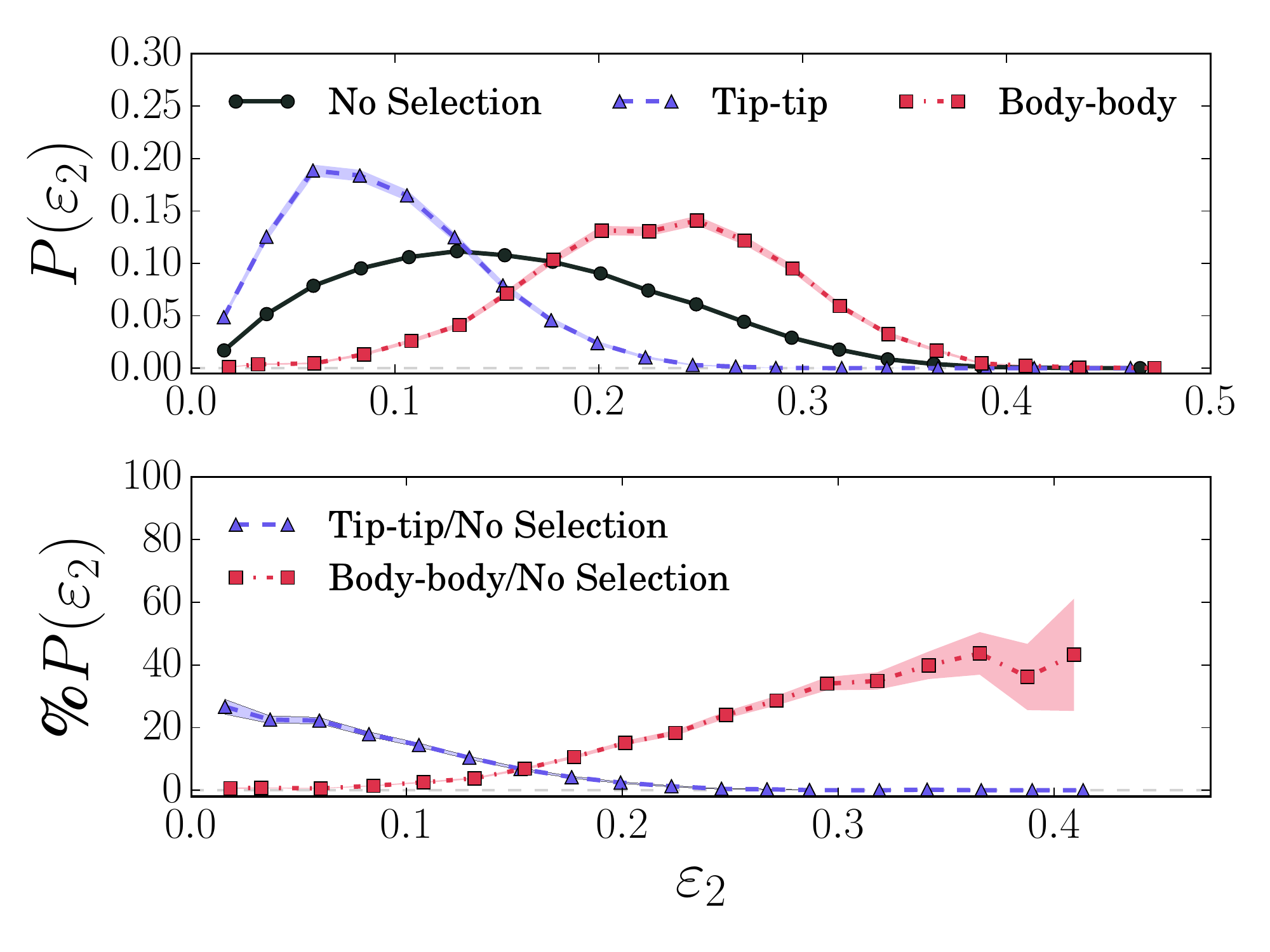}}
\end{tabular}
  \caption{Probability distribution for $dS/dy$ (left) and $\varepsilon_2$ (right) for different event classes within a sample of 1\% ZDC events.  The top panels show the distributions of tip-tip and body-body collisions scaled according to their contribution to the total population within the 1\% ZDC sample. The bottom panels show relative probabilities for tip-tip and body-body events among all events of a given $dS/dy$ (left) or $\varepsilon_2$ (right).}
  \label{fig:4}
\end{figure*}

In Fig.~\ref{fig:4} we plot the probability distributions for $\varepsilon_2$ and $dS/dy$. Using our collision definitions we can directly read off from the figure the likelihood of selecting a certain orientation based on a given eccentricity or multiplicity cut. We see in the left bottom panel that by cutting (within our 1\% ZDC sample) on events with large $dS/dy$ we can enrich the fraction of tip-tip events to about 50\%, whereas cutting on low $dS/dy$ enriches the fraction of body-body events, but never to more than about 20\%. The 20\% limit arises from admixtures from imperfectly aligned collisions that are not really "full overlap''.  The enrichment of tip-tip or body-body by varying $dS/dy$ relies on the assumed two-component nature of entropy production which also produced the knee structure discussed before. Indeed, selection efficiency of specific collision geometries by cutting on $dS/dy$ is model dependent.

We therefore consider ``event engineering'', i.e. selecting events by the magnitude of their $v_2$ flow vectors (for us, of the linearly related $\varepsilon_2$), shown in the right panels of Fig.~\ref{fig:4}.  Since tip-tip events have on average smaller ellipticities (see upper right panel), selecting events with small ellipticity (or, in experiment, small $v_2$) enriches the tip-tip fraction. However, in this way we will never reach more than about 25\% purity of the tip-tip sample. On the other hand, cutting the 1\% ZDC events on large $\varepsilon_2$ (or $v_2$) will enrich the sample in body-body events, with a purity that can reach about 40\% for the largest $\varepsilon_2$ values. While we have not yet been able to verify this with an actual cut on $v_2$, we expect this feature to survive the hydrodynamic evolution due to the almost perfect linearity between $\varepsilon_2$ and $v_2$.

\begin{figure}
  \centering
  \includegraphics[width = 0.7\textwidth]{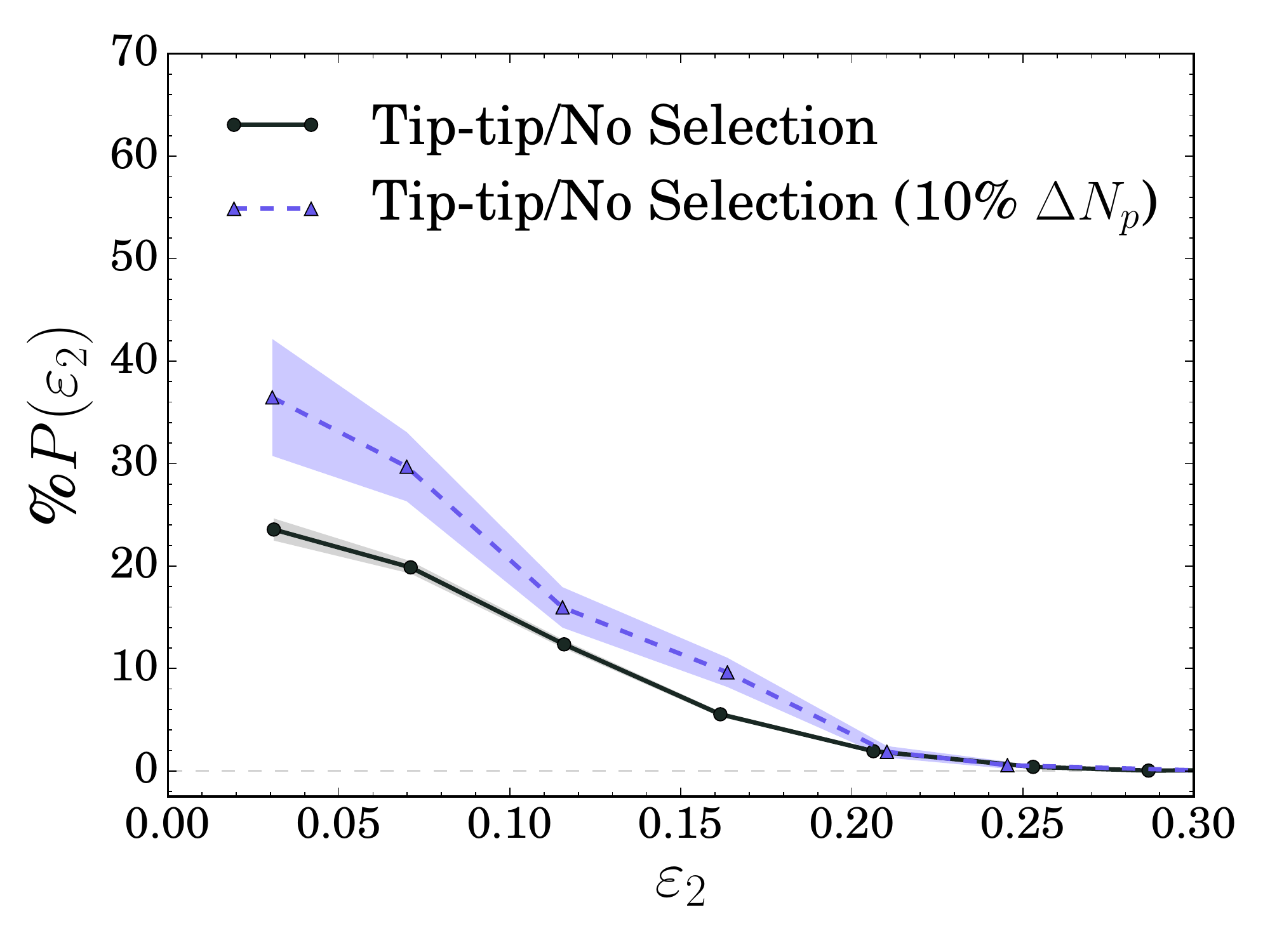}
  \caption{The black curve is the distribution of tip-tip events scaled according to their contribution to the total probability distribution for $\varepsilon_2$ as seen in the bottom right pannel in Fig.~\ref{fig:4}. The blue dashed curve shows the increased contribution of tip-tip collisions within the 10\% of events having the smallest difference in participants $\Delta N_\mathrm{part}$ (a proxy for ZDC correlation).}
\label{fig:5}
\end{figure}

The current ZDC cut strategy can be refined further to increase the probability of selecting tip-tip events.  Rather than looking at the ZDC signal in one of the two ZDC detectors or the sum of the ZDC signals in both detectors, we can look at the correlation of these two signals. Events with equal forward and backward ZDC signals (i.e. equal numbers of spectators from both nuclei) provide a better definition of the categories full overlap, tip-tip, and body-body than events with asymmetric ZDC signals where all spectators come from only one of the colliding nuclei. The difference in participants $\Delta N_\mathrm{part} = \left| N_\mathrm{part,1}-N_\mathrm{part,2} \right|$ quantifies the ZDC correlation in our model. Low values of $\Delta N_p$ correspond to the most correlated forward and backward ZDC signals. To demonstrate one application, we reconsider the events from the bottom right panel in Fig.~\ref{fig:4} and now select from the sample only events in the lowest 10\% of $\Delta N_\mathrm{part}$.  The selection on small values of $\Delta N_\mathrm{part}$ eliminates from the sample asymmetric configurations that we loosely describe as ``tip-body''.  Collisions of this type produce low values of $\varepsilon_2$ without the angular criteria necessary to be considered tip-tip and therefore dilute the contribution of the true tip-tip configurations at the lower range of $\varepsilon_2$. We show in Fig.~\ref{fig:5} that selecting the lowest 10\% of $\Delta N_\mathrm{part}$ increases the selection efficiency of $\varepsilon_2$ for tip-tip configurations by a factor of 1.4.  As a final comment, we point out that it might also be interesting to use ZDC correlations in the opposite way and to select and study events with asymmetric tip-body configurations.

\section{Conclusion}
\label{sec:4}
Within the two-component MC-Glauber model for initial energy production, the prolate deformation of the uranium nucleus was shown to generate a knee in the centrality dependence of the ellipticity of the initial temperature distribution.  The knee was seen to be preserved by hydrodynamic evolution, after which it manifests itself in the centrality dependence of $v_2$.  Such a knee structure is not seen in the STAR data. This rules out the two-component MC-Glauber model for initial energy and entropy production. An enrichment of tip-tip configurations by triggering only on high-multiplicity in the U+U collisions thus does not work.  

To increase the selection capability between different collision geometries, we impose combined cuts on initial conditions using the spectators (ZDC), $dS/dy$, and $\varepsilon_2$. For 1\% ZDC events, we found that we could enrich tip-tip collision geometries to about 50\% by cutting on high multiplicity within that sample, and body-body configurations to about 20\% purity by selecting low-multiplicity events. These numbers rely on the binary collision admixture in the two-component MC-Glauber model and are thus model-dependent. They do include effects from multiplicity fluctuations.

We also studied the efficiency of selecting different collision geometries by ``event engineering'', i.e. by cutting on $\varepsilon_2$ (by cutting on $v_2$ in the experiment). In this case events selected for high $\varepsilon_2$ can enrich body-body collisions to about 40\% purity while cutting on low $\varepsilon_2$ gives a tip-tip sample with about 25\% purity. The latter can be boosted to about 35\% purity by eliminating events with asymmetric ZDC signals. These results should not be sensitive to the binary collision admixture in the two-component MC-Glauber model and thus should be less model dependent.

\bigskip
\bigskip
{\noindent \bf Acknowledgments} \\
This work was supported in part by the U.S. Department of Energy, Office of Science, Office of Nuclear Physics under Award No. DE-SC0004286.

\end{document}